\documentclass[twocolumn,superscriptaddress,aps,pra,showpacs]{revtex4}
\usepackage{amsmath}
\usepackage{graphicx}
\usepackage{amsfonts}
\usepackage{subfigure}
\usepackage{amssymb}
\usepackage{hyperref}
\usepackage{color}
\usepackage{mathrsfs}
\usepackage{bbm}
\usepackage{times,txfonts}
\usepackage{isomath}
\usepackage{units}
\usepackage{nicefrac}
\usepackage{xcolor}

\begin{document}
\title{Continuous variable versus hybrid schemes for quantum teleportation of Gaussian states}

\author{Ioannis Kogias}
\email{john$_$k$_$423@yahoo.gr}
\affiliation{School of Mathematical Sciences, The University of Nottingham,
University Park, Nottingham NG7 2RD, United Kingdom}

\author{Sammy Ragy}
\email{pmxsr3@nottingham.ac.uk}
\affiliation{School of Mathematical Sciences, The University of Nottingham,
University Park, Nottingham NG7 2RD, United Kingdom}

\author{Gerardo Adesso}
\email{Gerardo.Adesso@nottingham.ac.uk}
\affiliation{School of Mathematical Sciences, The University of Nottingham,
University Park, Nottingham NG7 2RD, United Kingdom}

\begin{abstract}
In this paper, we examine and compare two fundamentally different teleportation schemes; the well-known continuous variable scheme of Vaidman, Braunstein and Kimble (VBK), and a recently proposed hybrid scheme by Andersen and Ralph (AR). We analyze the teleportation of ensembles of arbitrary pure single-mode Gaussian states using these schemes and see how they fare against the optimal measure-and-prepare strategies -- the benchmarks. In the VBK case, we allow for non-unit gain tuning and additionally consider a class of non-Gaussian resources in order to optimize performance. The results suggest that the AR scheme may likely be a more suitable candidate for beating the benchmarks in the teleportation of squeezing, capable of achieving this for moderate resources in comparison to the VBK scheme. Moreover, our quantification of resources, whereby different protocols are compared at fixed values of the entanglement entropy or the mean energy of the resource states, brings into question any advantage due to non-Gaussianity for quantum teleportation of Gaussian states.
\end{abstract}

\date{\today}

\pacs{03.67.Ac, 03.65.Ud, 03.67.Hk, 42.50.Ex}

\maketitle

\section{Introduction}

Quantum teleportation \cite{telep,vaidman,brakim} is a cornerstone of quantum information, simple enough to be taught in introductory-level quantum information courses, yet important enough to maintain a position at the forefront of contemporary research.
In practical terms, teleportation is an indispensable tool for the transmission of quantum information. This stands as one of the pillars of a networked system, along with storage and processing. In the past two decades there has been significant experimental progress in the field of teleportation, on a variety of different systems \cite{boschi,telezei,furuscience,teleatom1,teleatom2,memorypolzik, telepolzik,telesqz1,telesqz2,memorysqz1,memorysqz2,remotememory,memolaurat,naturusawa,fernnp13,memsaab, memory10,furuscience12,fernnp,zeil144}. An important class of these are continuous variable  systems, which range from atomic ensembles to optical modes and beyond \cite{book,brareview}.

One product of the focus on quantum teleportation has been the development of teleportation benchmarks \cite{brajmo,popqub,bancalnew,benchnew,benchqub,benchd,hammerer,owari,mariona,noi,peterb}. Put crudely, these benchmarks determine how good a teleportation-like procedure must be such that it could have been performed only with a shared entangled resource. Due to the relative difficulty of creating and maintaining long distance entanglement, these benchmarks are of practical interest as well as theoretical. For Gaussian states, which compose some of our most practical and popular continuous variable resources  (as well as including the set of all `classical' optical states \cite{hammerer}), general benchmarks for quantum teleportation have only very recently been derived \cite{benchnew}.

\begin{figure}[t]
	\includegraphics[width=7cm]{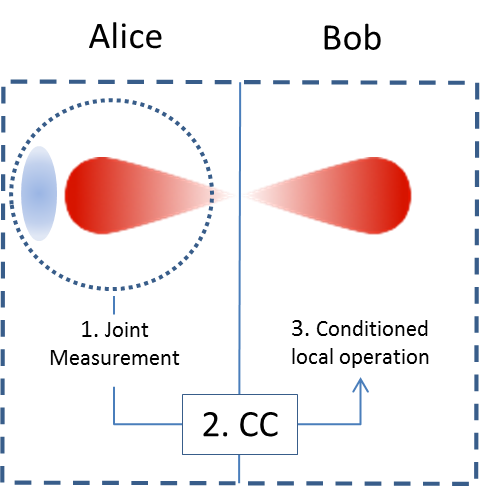}
	\caption{(Color online) A conceptual diagram for a general teleportation scheme. The leftmost (blue online) ellipse indicates the input state and the double cone (red online)  denotes the resource. The results of (1) a  joint measurement, performed by Alice, are (2) classically communicated (CC) to Bob, who performs (3) a local operation conditioned on the measurement result of Alice, in order to recreate the input state using his part of the resource.}
\label{TeleDiag}
\end{figure}

To clarify further, it is necessary to first decompose a quantum teleportation system into its essential components and procedures as in Fig.~\ref{TeleDiag}. We initialize the system by providing the state to be teleported (input) and a ``resource state''. Subsequently, Alice performs a joint measurement on the input and her part of the resource state and communicates the result to Bob, who performs a local operation on his state conditioned upon this measurement. The resource state, or set of resource states, which carries the entanglement shared between the two systems is what we consider to be the quantum part of the protocol. The classical communication conducted after Alice's measurement is by comparison very cheap, and thus we consider classical resources to be free, as is customary in quantum information resource theory.

To measure how `good' a teleportation is, for input and output states $\left|\psi  \right\rangle _{\text{in}}$ and $\hat{\rho}_{\text{out}}$ respectively, we use the fidelity \begin{equation}\mathcal{F}={}_{\text{in}}\left\langle \psi  \right|{{\hat \rho }_{\text{out}}}{\left| \psi  \right\rangle _{\text{in}}},\end{equation} for which $\mathcal{F}=1$ indicates a perfect teleportation \cite{uhlmann,jozsa}. A benchmark determines how large the average fidelity over a set of input states needs to be before it can be said with certainty that entanglement was necessary for the protocol used; that is, benchmarks set the limit on what a strategy can achieve using only local operations and classical communication.
In a sense, we might say that a quantum teleportation procedure is not \emph{truly} quantum unless it surpasses the optimal classical strategy in this regard: given some results from an unknown procedure, we can only definitively say that some entanglement was used if they exceed the benchmark.

In this paper we employ benchmarks recently derived by Chiribella and Adesso \cite{benchnew} in order to assess different teleportation schemes for general sets of single-mode Gaussian state inputs.  High-fidelity teleportation of Gaussian states is one essential ingredient for future realizations of quantum communication networks interfacing light and matter \cite{qinternet,hamrmp,cubic}, yet no effective scheme has been devised so far (to the best of our knowledge) to teleport effectively ensembles of squeezed states with limited resources.

We analyze the original single-mode Gaussian-state teleportation scheme, derived by Vaidman  \cite{vaidman}, and Braunstein and Kimble \cite{brakim} (VBK), in which a two-mode-squeezed vacuum state is used as the resource, and contrast this with a scheme recently introduced by Andersen and Ralph \cite{AR} (AR), where the quantum resource consists of $N$ two-qubit Bell states.

We find that the VBK teleportation is actually inferior to the AR teleportation within a particular realistic and important parameter range. This persists even when improvements to the VBK scheme are considered, such as gain tuning \cite{Bowen2003} and the possible introduction of sources of non-Gaussianity into the scheme. For a small amount of `resources' (to be quantified precisely in the following), the AR teleportation beats the VBK scheme in all considered variations, although in the presence of  larger amounts of  resources the advantage of the AR scheme fades away.
Notably, the VBK scheme requires in excess of $10$ dB of squeezing to exceed the benchmarks for teleportation of squeezed vacuum states without gain-tuning \cite{benchnew}. This value is teetering on the edge of the highest squeezing ever achieved in current optical experiments \cite{schnabel,schnabel13}, rendering untuned VBK teleportation incapable of beating the benchmarks even with state-of-the-art technology. Our analysis indicates that AR teleportation may provide a more viable candidate for this purpose.
There is, however, an important catch. A crucial difference between the two protocols is that the AR scheme is {\it probabilistic}, while the original VBK protocol is {\it deterministic}, or `unconditional' \cite{furuscience}. We dedicate ample discussion in the paper to address this point fairly.

The manuscript is organized as follows. In Sec.~\ref{secRev} we recall the continuous variable teleportation protocols we consider in this work, as well as the recently derived benchmarks for teleporting Gaussian states. In Sec.~\ref{Comparison} we lay down the terms of comparison we adopt for our analysis. In Sec.~\ref{SecRes} we present our detailed analysis on the performance of different schemes for teleporting ensembles of Gaussian states at fixed entanglement or energy. We draw our conclusions in Sec.~\ref{SecCon}.

\section{Continuous variable quantum teleportation schemes}\label{secRev}

\subsection{Vaidman-Braunstein-Kimble teleportation protocol}\label{VBK}

\begin{figure}[t]
\includegraphics[width=8cm]{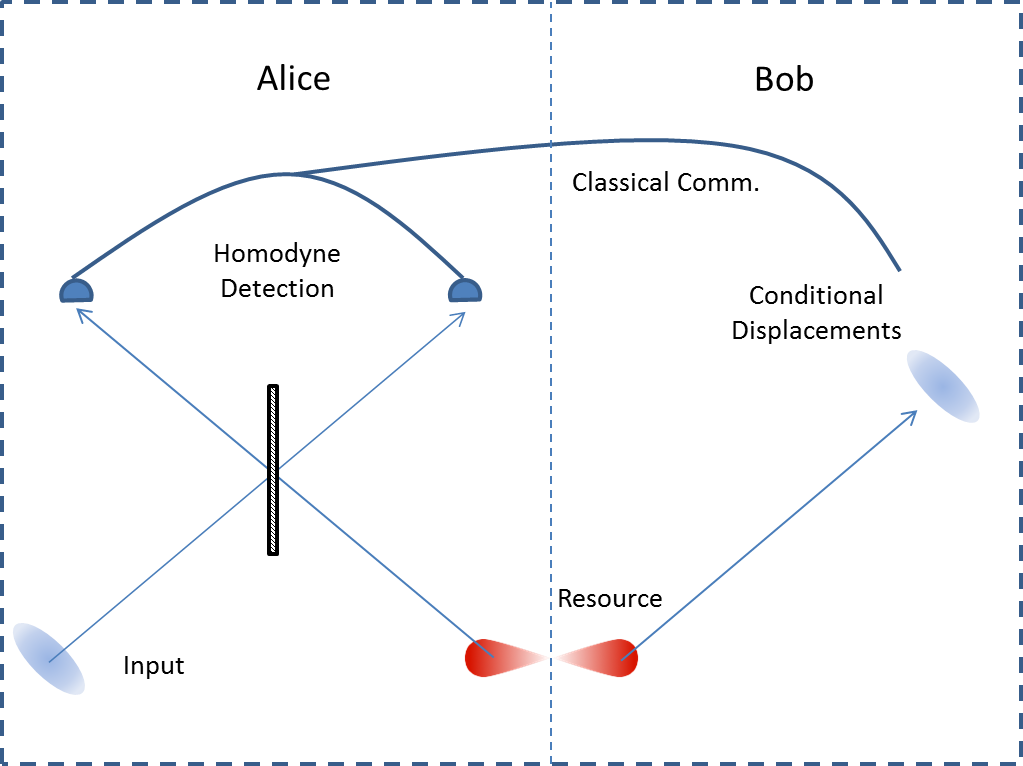}
\caption{(Color online) A schematic for the VBK teleportation scheme \cite{vaidman,brakim}. The shared resource state  is a two-mode entangled state.}
\label{BonkeyKong} \end{figure}

The first proposal of a teleportation protocol for continuous variable quantum states came from Vaidman and was shortly afterwards refined by Braunstein and Kimble \cite{vaidman,brakim} (VBK). In this protocol, two distant parties, Alice and Bob, share a two-mode continuous variable entangled state ${\hat \rho _{AB}}$ (resource) of modes $A$ and $B$ respectively. Alice wants to teleport an unknown quantum state ${\hat \rho _{\text{in}}}$ to Bob, and proceeds to do so by the protocol depicted in Fig.~ \ref{BonkeyKong}, with the following steps:

\begin{enumerate}
	\item Alice performs a 50/50 beam-splitting operation on the mode ${\hat \rho _{\text{in}}}$ that she wants to teleport and her share ${\hat \rho _{A}}$ of the two-mode entangled state ${\hat \rho _{AB}}$, yielding two output modes.
	\item Alice subsequently performs a homodyne measurement on each of the 
output modes, measuring two commuting position- and momentum-like observables $\hat{x}_+, \hat{p}_-$, and communicates the measurement outcomes  ${\tilde{x}_+},{\tilde{p}_-}$ to Bob via a classical channel.
	\item Bob uses Alice's measurement result to perform a suitable unitary displacement operation on his share ${\hat \rho _{B}}$ of the two-mode entangled state ${\hat \rho _{AB}}$, getting the output state  ${\hat \rho _{\text{out}}}$.
\end{enumerate}

Bob's output  ${\hat \rho _{\text{out}}}$ after the completion of the teleportation process is directly related to the entangled state ${\hat \rho _{AB}}$ and the input state ${\hat \rho _{\text{in}}}$. This relation has a simple expression in the characteristic function representation \cite{Marian2006,Dell'Anno2007},
\begin{equation} \label{1}
\begin{split}
{\chi _{\text{out}}}\left( \alpha \right)& = {\rm Tr}\left[ {{{\hat D}_{\text{out}}}\left(- \alpha \right){{\hat \rho }_{\text{out}}}} \right] \\
&  = {\chi _{\text{in}}}\left( {g\,\alpha} \right)\,{\chi _{AB}}\left( {g\,{\alpha^*},\alpha} \right),
\end{split}
\end{equation}
where $g$ is the so-called gain factor of the protocol \cite{Bowen2003}, $\hat D_k\left( \alpha  \right) = \exp [\alpha {{\hat a_k}^\dag } - {\alpha^*}\hat a_k]$ is the displacement operator acting on the mode $k$ with annihilation operator $\hat{a}_k$, and
\begin{gather} \label{2}
{\chi _{\text{in}}}\left( \alpha  \right) = {\rm Tr}\left[ {{{\hat D}_{\text{in}}}\left(- \alpha \right){{\hat \rho }_{\text{in}}}} \right], \\ \label{3}
 \,{\chi _{AB}}\left( {{\alpha_1},{\alpha_2}} \right) = {\rm Tr}\left[ {{{\hat D}_A}\left( {{ - \alpha_1}} \right){{\hat D}_B}\left( {{- \alpha_2}} \right){{\hat \rho }_{AB}}} \right],
\end{gather}
are the characteristic functions of the input state and the two-mode entangled states respectively.
The gain factor $g$ is chosen by Bob when he performs the displacement of his mode in \textit{Step 3}. In the ideal case where the shared entanglement between Alice and Bob is maximal (i.e. infinite), the teleportation performance is optimal  for gain $g=1$. However, in a realistic scenario of finite entanglement, the optimal choice of $g$ is not equal to 1. The fidelity $\cal F$ \cite{brajmo} can be computed by the above formalism with a formula, which for pure input states takes the form
\begin{equation} \label{4}
\begin{split}
\mathcal{F}_{VBK} & = {}_{\text{in}}\left\langle \psi  \right|{{\hat \rho }_{\text{out}}}{\left| \psi  \right\rangle _{\text{in}}} \\
& = \frac{1}{\pi }\int {{d^2}\alpha\,\,{\chi _{\text{in}}}\left( \alpha  \right)\,{\chi _{\text{out}}}\left( { - \alpha} \right)}.
\end{split}
\end{equation}
By using Eq. \eqref{1} we can express the fidelity solely w.r.t. the characteristic functions of the input and resource states,
\begin{equation}\label{5}
\mathcal{F}_{VBK} = \frac{1}{\pi }\int {{d^2}\alpha\,\,{\chi _{\text{in}}}\left( \alpha  \right)\,{\chi _{\text{in}}}\left( { - g\,\alpha} \right)\,{\chi _{AB}}\left( { - g\,{\alpha^*}, - \alpha} \right)}.
\end{equation}
For resource states $\hat{\rho}_{AB}$ with finite entanglement, one has $\mathcal{F} < 1$ strictly. Thus, a major contrast of this protocol with teleportation of finite-dimensional systems is that, even in principle, a perfect fidelity cannot be  achieved. Even worse, in practice, large amounts of entanglement cannot be achieved. In an attempt to overcome this difficulty, a new teleportation scheme has been recently proposed, which we will examine next.

\subsection{Andersen-Ralph teleportation protocol}

\begin{figure}[t]
\includegraphics[width=8.5cm]{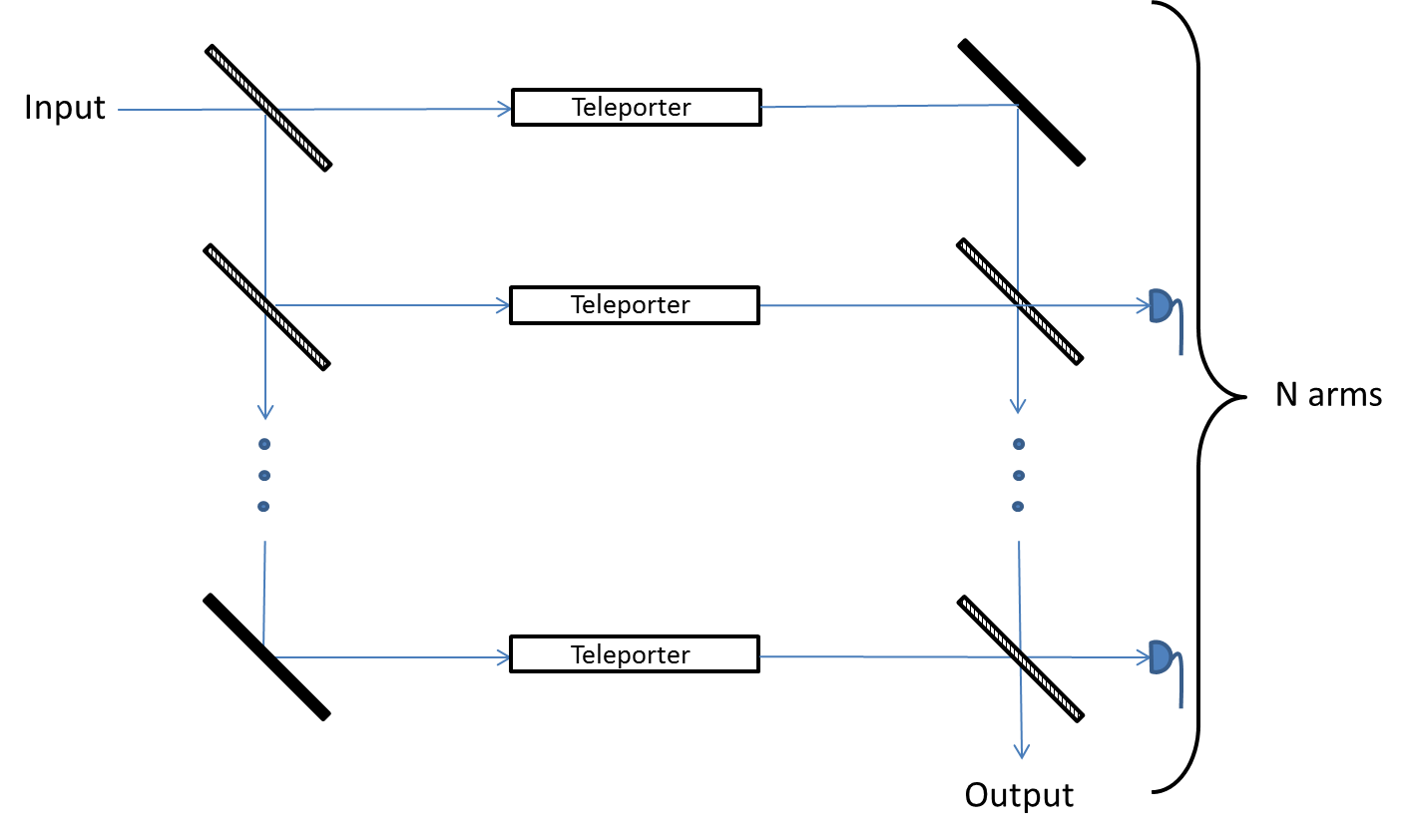}
\caption{(Color online) A schematic for the AR teleportation scheme \cite{AR}. The shared resources are $N$ two-qubit Bell states. Each teleporter is a typical qubit teleporter as originally introduced in \cite{telep}. The dark solid rectangles at the (bottom-left and top-right) corners indicate mirrors, and the other striped ones indicate beam splitters.}
\label{ARdiag} \end{figure}

The idea of the Andersen and Ralph (AR) scheme \cite{AR}, illustrated in Fig.~\ref{ARdiag}, is to remove the need for a single resource state with large entanglement, replacing it by multiple ones with lesser entanglement. This is done by splitting the input state using an $N$-splitter network to create $N$ identical modes (preferably with a vanishing probability of having more than one mean photon per mode). In the coherent state basis this global beam-splitter transformation of the input state takes the following form,
\begin{equation}\label{6}
\int {{d^2}\alpha\,} {\left\langle {\alpha}
 \mathrel{\left | {\vphantom {\alpha \psi }}
 \right. \kern-\nulldelimiterspace}
 {\psi } \right\rangle _{in}}\,\,{\left| \alpha \right\rangle} \to \int {{d^2}\alpha\,} {\left\langle {\alpha}
 \mathrel{\left | {\vphantom {\alpha \psi }}
 \right. \kern-\nulldelimiterspace}
 {\psi } \right\rangle _{in}}\,\,\left| {\frac{\alpha}{{\sqrt N }}} \right\rangle _{}^{ \otimes N},
\end{equation}
The $N$ split inputs are then truncated into states of the form $c_0|0\rangle +c_1 |1\rangle$ and can be separately teleported using $N$ maximally entangled two-qubit Bell states:
\begin{equation}\label{7}
{\left| \phi  \right\rangle _{AB}} = \frac{1}{{\sqrt 2 }}\left( {{{\left| {10} \right\rangle }_{AB}} + {{\left| {01} \right\rangle }_{AB}}} \right),
\end{equation}
where $\left| {0} \right\rangle$ and  $\left| {1} \right\rangle$ are the vacuum and one-photon states respectively.
At the output, the $N$ teleported modes are recombined in a similar beam-splitter network to produce the final output multiphoton state, which takes the form \cite{AR}
\begin{equation}\label{8}
{\left| \Psi  \right\rangle _{\text{out}}} = \frac{1}{{\sqrt {{P_{\text{suc}}(|\psi_{\text{in}}\rangle)}} }}\sum\limits_{k = 0}^N {{{\left\langle {k}
 \mathrel{\left | {\vphantom {k \psi }}
 \right. \kern-\nulldelimiterspace}
 {\psi } \right\rangle }_{\text{in}}}} \left( {\begin{array}{*{20}{c}}
   N  \\
   k  \\
\end{array}} \right)\frac{{k!}}{{{N^k}}}\,\,{\left| k \right\rangle _{\text{out}}},
\end{equation}
where the input-state dependent normalization constant $P_{\text{suc}}(|\psi_{\text{in}}\rangle)$ is defined as
\begin{equation}\label{10}
P_{\text{suc}}(|\psi_{\text{in}}\rangle) = {\sum\limits_{k = 0}^N {{{\left| {{{\left\langle {k}
 \mathrel{\left | {\vphantom {k \psi }}
 \right. \kern-\nulldelimiterspace}
 {\psi } \right\rangle }_{\text{in}}}} \right|}^2}\left( {\begin{array}{*{20}{c}}
   N  \\
   k  \\
\end{array}} \right)} ^2}\frac{{k{!^2}}}{{{N^{2k}}}}.
\end{equation}
The quality of the teleportation process will be quantified by the fidelity, which is found to be
\begin{equation}\label{9}
\mathcal{F}_{AR} = {\left| {{}_{\text{in}}{{\left\langle {\psi }
 \mathrel{\left | {\vphantom {\psi  \Psi }}
 \right. \kern-\nulldelimiterspace}
 {\Psi } \right\rangle }_{\text{out}}}} \right|^2} = \frac{1}{{{P_{\text{suc}}(|\psi_{\text{in}}\rangle)}}}{\left| {\sum\limits_{k = 0}^N {\left( {\begin{array}{*{20}{c}}
   N  \\
   k  \\
\end{array}} \right)\frac{{k!}}{{{N^k}}}{{\left| {{{\left\langle {k}
 \mathrel{\left | {\vphantom {k \psi }}
 \right. \kern-\nulldelimiterspace}
 {\psi } \right\rangle }_{\text{in}}}} \right|}^2}}} \right|^2}.
\end{equation}
In principle, this protocol allows large amounts of shared entanglement to be exploited by dividing it amongst the $N$ single-photon teleporters, removing the need for large two-mode squeezing as in the VBK protocol. However, the protocol is intrinsically \textit{probabilistic}, in that occasionally no output will be registered, for two reasons. The first is the truncation procedure: if large photon-number terms exist with significant probability in the state $|\psi\rangle_{\text{in}}$ then projecting onto the $\{|0\rangle\langle 0|, |1\rangle\langle 1|\}$ sector of the Fock space may have only a small chance of success. Secondly, to recombine the $N$ teleported modes, we demand all the photons to exit only one port, i.e. we wish to measure $|0\rangle$ in each of the detectors of Fig. \ref{ARdiag}, while in any other case the protocol fails. The overall probability of success of the AR scheme is none other than the aforementioned normalization factor $P_{\text{suc}}(|\psi_{\text{in}}\rangle)$, Eq.~(\ref{10}).

\subsection{Teleportation benchmarks} \label{Quantum benchmarks}
Benchmarks provide a fidelity threshold $\bar{\mathcal{F}}_c$, corresponding to the maximum average fidelity that can be achieved by classical measure and prepare schemes, without the two parties sharing any entangled resources, see e.g.~\cite{brajmo}. We consider in general {\it probabilistic} measure and prepare strategies, according to which we restrict our output to when we have a successful measurement and entirely discard and ignore the outputs for when we do not. Expressing this mathematically, we have \cite{giulionew,benchnew}
\begin{equation}\label{eqben}
\bar{\mathcal{F}}_c = \sum_{x\in X} \sum_{y\in Y_{\text{suc}}} p(x|suc)
\frac{\langle\psi_x|\hat{\Pi}_y|\psi_x \rangle}{\sum_{y'\in Y_{\text{suc}}}\langle\psi_x|\hat{\Pi}_{y'}|\psi_x \rangle}\langle\psi_x|\hat\rho_y|\psi_x \rangle
\end{equation}
Here, our measurement consists of the positive-operator-valued-measure elements $\lbrace\hat{\Pi}_y\rbrace$ and we discard all output results when $y \notin Y_{\text{suc}}$ where $Y_{\text{suc}}$ constitutes the set of what we consider to be favourable outcomes. Additionally, $p(x|\text{suc})$ denotes the probability that, given a successful outcome, the input state was $|\psi_x\rangle$ and finally, the term $\langle\psi_x|\hat\rho_y|\psi_x \rangle$ represents the corresponding fidelity where we prepare the state $\hat\rho_y$ conditioned on an output $y$.

To derive benchmarks, it is necessary to define a prior probability distribution (henceforth \textit{prior}), from which the input states to be teleported are drawn. This is also a realistic requirement (rather than always choosing a flat prior) since in a laboratory setting, constraints imposed by the apparatus, such as on the energy of producible states, will automatically impose a somehow nontrivial prior.

Estimating the best classical strategy is a hard problem, and only partial results were known for specific classes of input states (e.g. coherent states \cite{hammerer}). The general benchmark for teleporting arbitrary pure single-mode Gaussian states was only recently derived by Chiribella and Adesso \cite{benchnew}; the authors calculated the classical fidelity threshold for two classes of input single-mode states, namely undisplaced squeezed states, and general (displaced squeezed) pure Gaussian states.

\subsubsection{Benchmark for arbitrary squeezed vacuum states}
We consider an input ensemble containing  squeezed states,
\begin{equation}\label{11}
\left| \xi  \right\rangle  = \hat S\left( \xi  \right)\left| 0 \right\rangle, \,
\end{equation}
where $\hat S\left( \xi  \right) = \exp [ - \frac{\xi}{2} {{\hat a}^{\dag 2}} + \frac{{\xi ^*}}{2}{{\hat a}^2}]\,$ is the single-mode squeezing operator and $\,\xi  = s\,{e^{i\varphi }}$ is an arbitrary complex squeezing parameter. A state with complex squeezing $\xi$ is drawn from the input ensemble according to the prior
\begin{equation}\label{12}
p_\beta ^S\left( {s,\varphi } \right) = \frac{1}{{2\pi }}\frac{{\beta \sinh s}}{{{{\left( {\cosh s} \right)}^{\beta  + 1}}}},
\end{equation}
where ${\beta ^{ - 1}}$ adjusts the width of the squeezing distribution, while the phase $\varphi$ is uniformly distributed, yielding the $\frac{1}{2 \pi}$ prefactor. For a given $\beta$, the classical fidelity threshold is found to be,
\begin{equation}\label{13}
\bar {\mathcal F} _c^S\left( \beta  \right) = \frac{{1 + \beta }}{{2 + \beta }}.
\end{equation}
We see that even when Alice is completely ignorant about the squeezing of the state drawn, i.e. when $\beta  \to 0$, the fidelity achieved without any entanglement is $\frac{1}{2}$ \cite{benchnew}. This is analogous to the benchmark for non-squeezed, coherent input states with totally unknown displacement \cite{hammerer}.
\
\\

\subsubsection{Benchmark for general displaced squeezed Gaussian states}
A general pure, single-mode Gaussian state can be represented as a displaced squeezed state,
\begin{equation}\label{14}
\left| {\alpha ,\xi } \right\rangle  = \hat D\left( \alpha  \right)\hat S\left( \xi  \right)\left| 0 \right\rangle,
\end{equation}
where $\hat D\left( \alpha  \right)$ is the displacement operator and $\hat S\left( \xi  \right)$ the squeezing operator defined above. A state, with displacement amplitude $\alpha$ and complex squeezing $\xi$, is drawn from the input ensemble according to the probability distribution,
\begin{multline}\label{15}
p_{\lambda, \beta }^G\left( {\alpha ,s,\varphi} \right) = \frac{{\lambda \beta }}{{2{\pi ^2}}}\frac{{\sinh s}}{{{{\left( {\cosh s} \right)}^{\beta  + 2}}}} e^{{ - \lambda {{\left| \alpha  \right|}^2} + \lambda {\mathop{\rm Re}\nolimits} \left( {{e^{ - i\varphi }}{\alpha ^2}} \right)\tanh s}},
\end{multline}
where ${\beta ^{ - 1}, \lambda ^{ - 1}}$ adjust the widths of the squeezing and displacement distributions, respectively. Note that this distribution correctly reproduces the probability distribution \eqref{12}  for squeezed-only states, $\int {{d^2}\alpha} \,\,p_{\lambda ,\beta }^G\left( {\alpha ,s,\varphi } \right) = p_\beta ^S\left( {s,\varphi } \right).$
For given $\beta, \lambda$, the classical fidelity threshold for this ensemble is found to be,
\begin{equation}\label{16}
\bar {\cal F }_c^G\left( {\lambda,\beta } \right) = \left( {\frac{{1 + \lambda }}{{2 + \lambda }}} \right) \left( {\frac{{1 + \beta }}{{2 + \beta }}} \right).
\end{equation}
When Alice is completely ignorant of both the displacement and the squeezing of the state drawn, i.e. $\lambda  \to 0$ and $\beta  \to 0$, the best achievable fidelity without use of any entanglement is $\frac{1}{4}$ \cite{benchnew}.

\section{Comparison of the teleportation protocols: Quantifying resources}\label{Comparison}
A vital topic to tackle for the understanding of this paper, and to facilitate fair comparison of teleportation schemes in general, is how to quantify {\it resources}. For a quantum teleportation scheme, it is customary to consider the resource to be the entangled state shared. We have then some freedom on what property of the resource state to choose for quantification and comparison. For the purposes of this paper, we choose two quantifiers as resources: the mean energy and the entanglement degree of the shared entangled state, and we perform independent comparisons of different schemes for given values of each.

Henceforth, \textit{entanglement} is synonymous with entropy of entanglement, defined for a pure resource state $\hat\rho_{AB}=|\phi\rangle_{AB}\langle\phi|$ as the von Neumann entropy, \begin{equation}S(\hat\rho_{A}) = - {\rm Tr}\left[ {{{\hat \rho }_A}{{\log }_2}{{\hat \rho }_A}} \right],\end{equation} of the reduced state $\hat\rho_A=\text{Tr}(\hat\rho_{{AB}})$. Additionally, \textit{energy} is defined by the total mean photon number in the modes $A$ and $B$, \begin{equation}{E_\phi} = \left\langle {{{\hat a}^\dag }_A{{\hat a}_A}} \right\rangle  + \left\langle {{{\hat a}^\dag }_B{{\hat a}_B}} \right\rangle,\end{equation}
where $\hat{a}_{A,B}$ refers to the bosonic annihilation operator for mode $A$, $B$ respectively.

These quantities are fairly straightforward to employ for comparing deterministic teleportation protocols; however, it is not immediately obvious how to compare probabilistic teleportations with differing success probabilities. In practice, furthermore, the resources truly utilized in any teleportation experiment are much more complicated than just these two quantities: everything from the energy used to power the equipment, to the manpower required to build it can be considered a resource if we wish to be omnicomprehensive in our definitions. While we certainly shall not explicitly consider these factors, they do implicitly impact in a very significant way to how we compare probabilistic teleportation schemes.

To this effect, we consider two possible interpretations for how we consider resources. The first interpretation counts the average resources required to achieve the teleportation of a state: we refer to this as the \emph{naive picture}, since it only counts the units of energy or entanglement, with no other weighting. For example, a two-arm AR scheme with a $50\%$ probability of success would require 2 runs of 2 ebits and thus use 4 ebits of entanglement per successful teleportation on average.
However, this interpretation is not suitable for practical comparisons: it builds a false equivalence between, for example, one usage of a 4-arm AR interferometer and two usages of a 2-arm interferometer. In practice, a 4-arm interferometer would be comparatively much more costly to assemble. Similarly, 4 ebits in the VBK scheme correspond to $13.7$ dB of entanglement, and the current experimental limit is about $10$ dB \cite{schnabel,schnabel13}, whereas 2 ebits correspond to a value of $7.7$ dB, which is fairly achievable; in this sense, two uses of a 2 ebit scheme are not comparable to one use of a 4 ebit scheme, in general, due primarily to the technological limitations of creating the extra entanglement.

We therefore adopt a \emph{pragmatic picture}, whereby we attempt to account for the realistic limitations on teleportation schemes.
To do this we first assume that producing the input states for teleportation is effectively free. As such, nothing important is lost on a failed teleportation attempt: this assumption is consistent with the formulation of the benchmarks, for which we freely discard states upon unsuccessful measurement outcomes. Indeed, even for deterministic schemes, thousands of (normally unaccounted for) independent runs are in practice repeated in the lab for a given input state, in order to perform state tomography on the output for experimental determination of the teleportation fidelity. In essence, building a teleportation setup is costly (in terms of acquiring a certain entanglement source, for instance), while running it repeatedly is assumed to be cheap in comparison. Furthermore, as we have been assuming all along, the classical communication required for teleportation is so cheap in comparison to entanglement that it can be neglected in our quantitative comparison.
For all of the above, in the pragmatic approach we choose to ultimately ignore the probability of success for a scheme (or equivalently the number of runs required to achieve a certain fidelity), and merely compare the number of ebits or units of energy (e.g.~photons, phonons) utilized in individual runs, whether successful or not. While a fully objective comparison of different schemes is perhaps not possible in principle, we believe this approach is fair and sufficient.

With this point of view in mind, it can be shown \cite{giulionew} that a general (possibly probabilistic) {\it quantum} teleportation protocol yields an average fidelity over a certain input ensemble given by the formula
\begin{equation}\label{DoNotLabelWithNumbersGiannis}
\bar{\mathcal{F}}_q = \sum_{x\in X} \sum_{y\in Y_{\text{suc}}} p(x|\text{suc})
\frac{\langle\Psi_{x,r}|\hat{\Pi}_y|\Psi_{x,r} \rangle}{\sum_{y'\in Y_{\text{suc}}}\langle\Psi_{x,r}|\hat{\Pi}_{y'}|\Psi_{x,r} \rangle}\langle\psi_x|\hat\rho_y|\psi_x \rangle.
\end{equation}
Note how this only differs from the equation for the classical benchmark (\ref{eqben}) in that, in the quantum case, we do not consider a measurement directly upon the input state, but rather upon the joint state $|\Psi_{x,r}\rangle=|\psi_x\rangle\otimes|\phi_r\rangle$, where $|\phi_r\rangle \equiv |\phi\rangle_{AB}$ refers to the shared resource state.

To summarize, then, we simply define our resources by the value of entanglement (in ebits) or energy (in units) of $|\phi\rangle_{AB}$ irrespective of any other factor.

\subsection{Resources for the AR scheme}
In the case of the AR scheme, the natural choice for the resource states is given by the maximally entangled two-photon Bell states, e.g. ${\left| \phi  \right\rangle _{AB}} = \frac{1}{{\sqrt 2 }}\left( {{{\left| {10} \right\rangle }_{AB}} + {{\left| {01} \right\rangle }_{AB}}} \right)$, since with these states we can achieve perfect teleportation in the $\left\{ {\left| 0 \right\rangle ,\left| 1 \right\rangle } \right\}$ subspace \cite{telep}. As the von Neumann entropy of a Bell state amounts to $1$ ebit,  for an $N$-arm set up with $N$ Bell states the total entanglement resource is given straightforwardly (exploiting additivity of the von Neumann entropy) by
\begin{equation}\label{16b}
S_{AR}\left( {{{\left| \phi  \right\rangle }_{AB}}{{\left\langle \phi  \right|}^{ \otimes N}}} \right) = N \, {\rm{{ebits}}}.
\end{equation}
Similarly, the energy of the resource states ${{{\left| \phi  \right\rangle }_{\text{AR}}}{{\left\langle \phi  \right|}^{ \otimes N}}}$ is the sum of energies for each ${{{\left| \phi  \right\rangle }_{AB}}\left\langle \phi  \right|}$,
\begin{equation}\label{16c}
E_{\text{AR}}\left( {{{\left| \phi  \right\rangle }_{AB}}{{\left\langle \phi  \right|}^{ \otimes N}}} \right) = N\,\,{\rm{units}}{\rm{.}}
\end{equation}

\subsection{Resources for the VBK scheme}
In the VBK scheme we will consider shared entangled states which belong to a general non-Gaussian class encompassing so-called `squeezed Bell-like states', first studied by Dell'Anno {\it et al.} \cite{Dell'Anno2007},
\begin{equation}\label{17}
{\left| {{\phi _{SB}}} \right\rangle _{AB}} = {{\hat S}_{AB}}\left( \zeta  \right)\left[ {\cos \delta {{\left| {0,0} \right\rangle }_{AB}} + {e^{i\theta }}\sin \delta {{\left| {1,1} \right\rangle }_{AB}}} \right],
\end{equation}
where \begin{equation}{{\hat S}_{AB}}\left( \zeta  \right) = \exp [ - \zeta \hat a_A^\dag \hat a_B^\dag  + {\zeta ^*}\hat a_A^{}\hat a_B^{}]\end{equation} is the two-mode squeezing operator with complex squeezing $\zeta  = r\,{e^{i\varphi }}$ and ${\left| {n,m} \right\rangle _{AB}} = {\left| n \right\rangle _A} \otimes {\left| m \right\rangle _B}$ is a two-mode Fock state.

For $\delta=k \pi$ ($k \in \mathbb{Z})$ we get the well-known two-mode squeezed vacuum (TMSV) state, \begin{equation}\label{tmsv}
{{\hat S}_{AB}}\left( \zeta  \right){\left| {0,0} \right\rangle _{AB}},\end{equation} with squeezing $r$, that is, the paradigmatic Gaussian entangled resource state. For other values of $\delta$, we get non-Gaussian contributions, and we deem it interesting to investigate whether such non-Gaussianity provides an advantage over the use of conventional TMSV states \cite{Dell'Anno2007,Dell'Anno2010}, under the terms of comparison defined above.

In the characteristic function representation the state ${\left| {{\phi _{SB}}} \right\rangle _{AB}}$ has the form
\begin{multline}\label{17a}
{\chi _{SB}}\left( {{\alpha _1},{\alpha _2}} \right) = {e^{ - \frac{{{{\left| {{\xi _1}} \right|}^2} + {{\left| {{\xi _2}} \right|}^2}}}{2}}}\left[ {{\sin \delta \cos \delta   \left( {{e^{i\theta }}\xi _1^*\xi _2^* + {e^{ - i\theta }}\xi _1^{}\xi _2^{}} \right)  }} \right. \\
\left. { + {{\sin }^2}\delta \left( {1 - {{\left| {{\xi _1}} \right|}^2}} \right)\left( {1 - {{\left| {{\xi _2}} \right|}^2}} \right) + {{\cos }^2}\delta  } \right],
\end{multline}
where ${\xi _i} = {\alpha _i}\cosh r + {\alpha _j}{e^{i\varphi }}\sinh r,\,\, \left( i,\,j=1,\,2;\, i\neq j\right)$.

The entanglement ${S_{VBK}}\left( {r,\phi ,\delta ,\theta } \right)$ of squeezed Bell-like states can be expressed as a rather long formula \cite{Dell'Anno2007} which we omit here, limiting ourselves to note that it depends nontrivially on both the complex squeezing $\zeta$ and on the non-Gaussian mixing parameter $\delta$ and phase $\theta$.

The mean energy of these states has a  more concise form,
\begin{multline}\label{23}
E_{\text{VBK}}\left( {r,\varphi ,\delta ,\theta } \right)  = \left\langle {\hat a_A^\dag \hat a_A^{}} \right\rangle  + \left\langle {\hat a_B^\dag \hat a_B^{}} \right\rangle  \\
 = 2{\sinh ^2}r\left( {1 + {{\sin }^2}\delta } \right) + 2{\sin ^2}\delta \,{\cosh ^2}r \\
 - \sin 2\delta \,\sinh 2r\,\cos \left( {\theta  - \varphi } \right).
\end{multline}

\section{Results}\label{SecRes}
For accurate comparison to the benchmarks \cite{benchnew}, we must consider states drawn from the general class of pure Gaussian states $|\alpha,\xi\rangle$ of Eq. \eqref{14} with probabilities given by the same priors $p_{\lambda ,\beta }^G\left( {\alpha ,s,\varphi} \right)$ or $p_{\beta }^S( {s,\varphi})$ as used to derive the benchmarks.

We then find the average fidelity for general input states drawn from a prior characterized by widths $\lambda^{-1}$ and $\beta^{-1}$ for a scheme with resources (entanglement or energy) of value $N$ to be
\begin{equation}\label{MeanFidVBK}
{{{\bar {\cal F}}_{\text{VBK}} }}\left( {\lambda ,\,\beta ,\,N} \right) = \int {{d^2}\alpha\,d\varphi \,ds\,p_{\lambda ,\beta }^G\left( {\alpha ,s,\varphi } \right)} \,{ {\mathcal{F}_{\text{VBK}}}}\left( {\alpha,s,\varphi ;N} \right),
\end{equation}
for the deterministic VBK scheme, and
\begin{eqnarray}\label{MeanFidAR}
&&{\bar {\cal F} _{\text{AR}}}\left( {\lambda ,\,\beta ,\,N} \right) \\ &&= \frac{{\int {{d^2}\alpha \,d\varphi \,ds\,p_{\lambda ,\beta }^G\left( {\alpha ,s,\varphi } \right){P_{\text{suc}}}\left( {\alpha ,s,\varphi } \right)} \,{\cal F}_{\text{AR}}\left( {\alpha ,s,\varphi ;N} \right)}}{{\int {{d^2}\alpha \,d\varphi \,ds\,p_{\lambda ,\beta }^G\left( {\alpha ,s,\varphi } \right){P_{\text{suc}}}\left( {\alpha ,s,\varphi } \right)} \,}},\nonumber
\end{eqnarray}
for the probabilistic AR scheme, in accordance with Eq. \eqref{DoNotLabelWithNumbersGiannis}.

\begin{figure*}[t]
\subfigure[]{\label{Figure1}\includegraphics[height=5.5 cm]{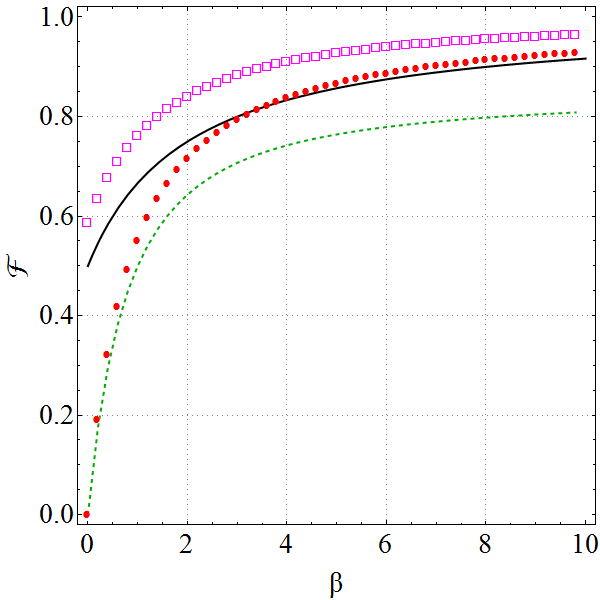}}\hspace*{.2cm}
\subfigure[]{\label{Figure2}\includegraphics[height=5.5 cm]{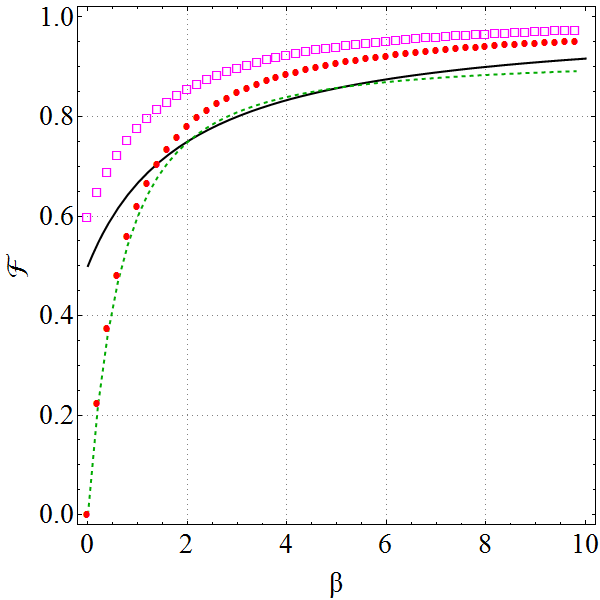}}\hspace*{.2cm}
\subfigure[]{\label{Figure3}\includegraphics[height=5.5 cm]{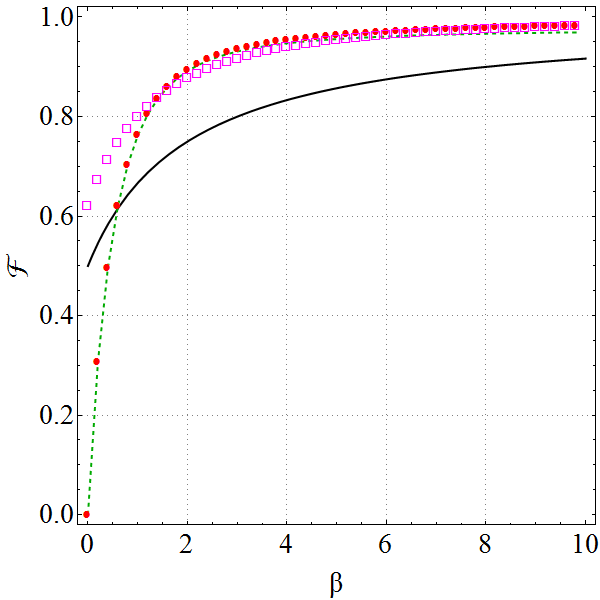}}
\caption{(Color online) Average fidelity of teleportation $\bar{\cal F}$ for the input set of single-mode squeezed states with prior $p_\beta ^S$, plotted as a function of the inverse width $\beta$, for different amounts of shared entanglement: (a) $S=2$ ebits, (b) $S=3$ ebits and (c) $S=5$ ebits. The comparison is between the AR scheme (magenta open squares), the VBK scheme optimized over all squeezed Bell-like resource states with unit gain (green dashed curve), the gain-tuned VBK scheme optimized over all squeezed Bell-like resource states, amounting to the gain-tuned VBK scheme using TMSV resource states (red filled circles), and the benchmark (black solid line).}
\label{Figures123}
\end{figure*}

Both fidelities reduce to the mean fidelity for squeezed-only states upon setting $\alpha=0$ and substituting the appropriate prior $p_{\beta }^S$ in place of $p_{\lambda ,\beta }^G$ (or, equivalently, taking the limit $\lambda \rightarrow \infty$ in the formulas above).
\subsection{Comparison I: Fixed entanglement entropy} \label{Comparison I}
We will study three different cases, when
\begin{equation}\label{20}
{S_{\text{AR}}}\left( {{{\left| \phi  \right\rangle }_{AB}}{{\left\langle \phi  \right|}^{ \otimes N}}} \right) = {S_{\text{VBK}}}\left( {r,\phi ,\delta ,\theta } \right) = 2,\,\,3,\, \rm{and}\,\, 5\,\,\,{\rm{ebits}}.
\end{equation}
For the AR scheme, this simply corresponds to considering $N=2,\,3\,\, \rm{and}\,\,5$ branches in the $N$-splitter, respectively. The teleportation fidelity of a general pure Gaussian input, $\left| \psi  \right\rangle_{\text{in}} = \left|\alpha,\, \xi  \right\rangle$, using Eq. \eqref{9}, is
\begin{equation}\label{21}
{\mathcal{F}_{\text{AR}}}\left( {\alpha,\,s,\varphi;N} \right)
= \frac{1}{{{P_{\text{suc}}}}}{\left|\sum\limits_{k = 0}^N {\left( {\begin{array}{*{20}{c}}
   N  \\
   k  \\
\end{array}} \right)\frac{{k!}}{{{N^k}}}{{\left| {{{\left\langle {k}
 \mathrel{\left | {\vphantom {k \psi }}
 \right. \kern-\nulldelimiterspace}
 {\alpha,\, \xi } \right\rangle }}} \right|}^2}} \right|^2},
\end{equation}
which can then be substituted into Eq. \eqref{MeanFidAR} to find the mean fidelity.

For the VBK scheme, from Eq. \eqref{5}, we see that the fidelity for teleporting a particular displaced squeezed state with characteristic function ${\chi _{\alpha,s,\varphi}}\left( \gamma  \right)$, via a two-mode squeezed Bell-like shared state, ${\chi _{SB}}\left( {{\gamma_A},{\gamma_B}} \right)$, is given by
\begin{multline}
{{\cal F}_{\text{VBK}}}\left( {\alpha,\,s,\varphi ;\,\,r,\phi ,\delta ,\theta ;\,\,g} \right) \\ = \frac{1}{\pi }\int {{d^2}\gamma \,\,{\chi _{\alpha,s,\varphi}}\left( \gamma  \right)\,}
 {\chi _{\alpha,s,\varphi}}\left( { - \gamma } \right)\,{\chi _{\text{SB}}}\left( { - g\,{\gamma ^*}, - \gamma } \right).
\end{multline}
This formula can be analytically evaluated for non-unit gain $g$, but the explicit expression is too long and cumbersome to be reported here.

Given the dependence of ${S_{\text{VBK}}}\left( {r,\phi ,\delta ,\theta } \right)$ on four different parameters, there is a manifold of states associated with any fixed value of entanglement, which can be found by numerically solving for each case of $N=2,\,\,3,\,\,5\,\,\,{\rm{ebits}}$. The optimal resource and best strategy can then obtained by optimizing the average fidelity, Eq.~(\ref{MeanFidVBK}), over the set of resource states with a given entanglement constraint $S=N$, and additionally optimizing over the gain $0 \leq g \leq 1$. This results in the optimal VBK average fidelity $\bar  {\mathcal{F}} _{\text{VBK}}^{\text{opt}}\left( {\lambda,\,\beta ,\,\,N} \right)$ given $N$ ebits of entanglement available in the form of squeezed Bell-like states.

In what follows, we compare the  average fidelities of the two teleportation schemes, ${\bar  {\mathcal{F}} _{\text{AR}}}\left(\lambda,\, \beta,\,N \right)$ and $\bar  {\mathcal{F}} _{\text{VBK}}^{opt}\left( {\lambda,\,\beta ,\,\,N} \right)$, as we vary the prior distribution parameters $\lambda$ and $\beta$.

\subsubsection{Results for squeezed states}
We begin by comparing the averaged fidelities ${{{\bar {\cal F}} }}_{\text{AR}}$ and ${{{\bar {\cal F}} }}_{\text{VBK}}$ as well as the corresponding benchmark $\bar{\cal F}_c^S$, for the case of teleporting squeezed states with zero displacement.

The first important result is depicted in Fig. \ref{Figure1}, where we set the entanglement resource value at $S=N=2$ ebits, for various values of $\beta$. The AR scheme manages to always beat the benchmark for every $\beta$, in sharp contrast to the VBK scheme, even for $\beta \to 0$. In this limit, which corresponds to completely unknown squeezing, the VBK teleportation scheme achieves negligible average fidelity, while both the AR scheme and the benchmark tend to finite values, $\bar {\mathcal{F}} _{\text{AR}} \to 0.58$  and $\bar {\mathcal{F}} _c^S \to 0.5$ respectively.  Even taking into account gain tuning, the optimized VBK scheme can just barely surpass the benchmark at large values $\beta$, does not look especially robust against possible experimental deficiencies. A conclusive experimental demonstration of quantum teleportation of an ensemble of squeezed states (with unknown squeezing) achieving fidelities superior to what is classically possible has yet to be achieved,
 and the present results indicate that the AR scheme may be a more viable candidate for this than the VBK scheme. The fact that only two branches are needed for such a demonstration, makes the scheme experimentally appealing with current technology. Clearly, the probabilistic nature of the AR scheme is a major factor behind its enhanced performance; such a scheme is indeed more likely to reject states which cannot be faithfully transmitted (i.e. high energy input states), and thus it compares favourably to the benchmark even in the limit  $\beta\to 0$. The VBK scheme on the other hand teleports the high energy states with vanishing fidelity, reducing the average fidelity to zero for very broad ensembles.

As we increase the entanglement entropy of the shared resource states to $S=3$ ebits, see Fig.~\ref{Figure2}, we find that the AR scheme is still superior, but now the VBK scheme clearly violates the benchmark for input ensembles of inverse width $\beta \geq 1.58$. For even greater entanglement of $S=5$ ebits,  Fig.~\ref{Figure3}, the VBK scheme manages to attain comparable performances to the AR one at large enough $\beta$, while the limit $\beta \to 0$ remains problematic. This level of shared resources is, however, unrealistic: state-of-the-art technologies achieve $10$ dB of optical squeezing \cite{schnabel,schnabel13} which is equivalent to only $2.77$ ebits of entanglement.

Another interesting result has to do with the performance of the squeezed Bell-like resource states for the VBK scheme. In \cite{Dell'Anno2007,Dell'Anno2010}, Dell'Anno {\it et al.} showed that, at fixed squeezing degree $r$, non-Gaussian squeezed Bell-like states (i.e., with $\delta \neq 0$) resulted in significant advantage in the teleportation fidelity of single coherent or squeezed states, compared to just using the corresponding Gaussian TMSV with the same $r$ (given by $\delta=0$). The authors thus concluded that non-Gaussianity in the resource state can significantly improve teleportation performance.

Our results show, however, that such a conclusion is strongly dependent on the terms of comparison. When making the comparison at fixed entanglement entropy, rather than at fixed squeezing degree, we found in all considered cases that, within the general squeezed Bell-like class, the optimal resource state for teleportation of input ensembles of Gaussian states via the gain-optimized VBK scheme actually {\it does} always reduce to the TMSV. In this respect, therefore,  non-Gaussianity is not advantageous for the considered task. One may contend that the advantage observed by Dell'Anno {\it et al.} was more properly a consequence of the extra entanglement present in the resource (compared to the TMSV at fixed $r$) and not traceable directly to the non-Gaussian nature of the employed states.

\subsubsection{Results for general displaced squeezed states}

\begin{figure*}[t]
\subfigure[]{\label{Figure4}\includegraphics[height=5.3cm]{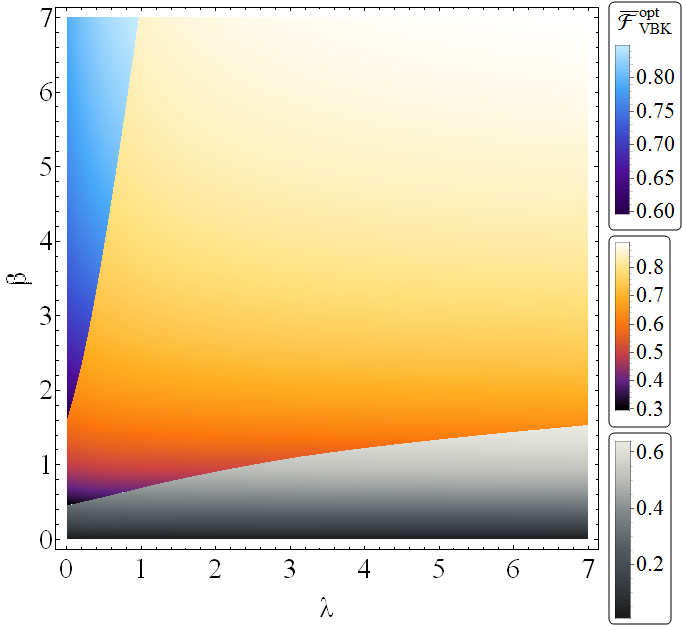}}\,
\subfigure[]{\label{Figure5}\includegraphics[height=5.3cm]{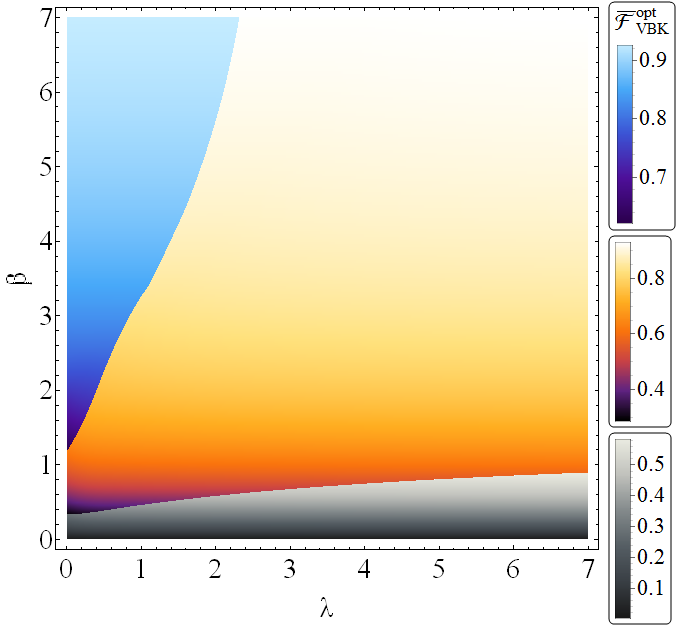}}\,
\subfigure[]{\label{Figure6}\includegraphics[height=5.3cm]{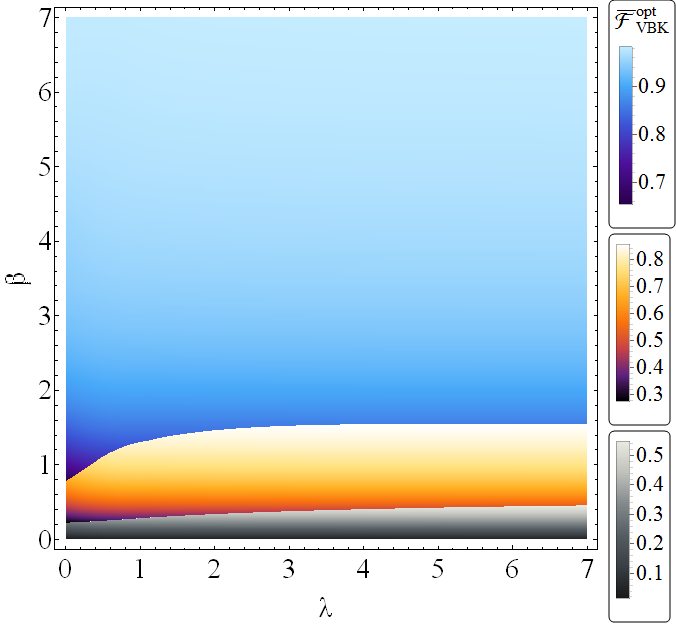}}
\caption{(Color online) Contour plots of the average teleportation fidelity $\bar  {\mathcal{F}} _{\text{VBK}}^{\text{opt}}$ for the input set of arbitrary displaced squeezed Gaussian states $|\alpha,\xi\rangle$ distributed according to the prior $p_{\lambda ,\beta }^G$, for the gain-optimized VBK scheme,   as a function of the inverse widths $\lambda$, $\beta$, at different fixed amounts of shared entanglement: (a) $S=2$ ebits, (b) $S=3$ ebits and (c) $S=5$ ebits. From top-left to bottom-right, the three shaded areas in each figure denote, respectively, the region where the VBK scheme has superior performance compared to both the AR scheme and the benchmark (sea colors online), the region where the VBK scheme is inferior to the AR one but still beats the benchmark (solar colors online) and the region where the VBK protocol yields a fidelity below the benchmark (grayscale colors online). The average fidelity of the AR protocol (not depicted) is found to always beat the benchmark for every value of the parameters $\lambda$, $\beta$.}
\label{Figures456}
\end{figure*}

We will now discuss the results for the most general set of pure single-mode Gaussian input states, namely the displaced, squeezed vacuum states. In Fig.~\ref{Figure4} we report the case of $S=2$ ebits of shared entanglement. As in the previous case of squeezed-only states, the AR scheme beats the benchmark for all values of the parameters $\beta, \ \lambda$. On the other hand, it no longer stands so dominant over the VBK scheme;  while for small $\beta$ and large $\lambda$ the AR scheme is still superior, as we increase $\beta$ and reduce $\lambda$ the optimized VBK scheme manages to achieve the best fidelity overall. This relates to the well-known result that the VBK scheme is exceptionally good, by construction, at teleporting displaced states (and in fact, despite being deterministic, always beats the benchmark for teleporting coherent states \cite{hammerer,telepoppy}). As we increase the shared entanglement to $S=3$ ebits, we see in Fig.~\ref{Figure5} that the dominance of the  AR scheme gets confined to the region of larger $\lambda$ and smaller $\beta$, while for the instance of even larger entanglement, $S=5$ ebits of Fig. \ref{Figure4}, the VBK protocol wins the comparison in almost the whole parameter region except for small $\beta$.

As in the previous subsection, we found again that non-Gaussianity in the shared squeezed Bell-like states yields no advantage in the VBK average teleportation fidelity over the conventional use of TMSV resources. Even in the present more general case of displaced squeezed input states, the fidelity depicted in Fig.~\ref{Figures456} corresponds in fact to the optimal choice given by the use of a TMSV resource state.


\subsection{Comparison II: Fixed mean energy}
In this section we will compare the two schemes by constraining the energy of their resource states, i.e. by keeping fixed the mean photon number at $E=N=2,\,\,3,\,\,5\,\,\,{\rm{units}}$, instead of the entanglement entropy which we considered previously.

As previously observed, the energy used in the AR scheme, $E_{\text{AR}}\left( {{{\left| \phi  \right\rangle }_{AB}}{{\left\langle \phi  \right|}^{ \otimes N}}} \right) = N\,\,{\rm{units}}$, is determined by the number of branches in exactly the same way as the entanglement entropy is: each branch corresponds to one ebit of entanglement and one unit of energy. Thus the fidelity of the scheme will still be given by \eqref{MeanFidAR}, and the performance of the scheme is the same as for the fixed entanglement case.

For the VBK scheme, however, the mean energy has a different dependence on the resource state parameters; to identify the optimal resources in the manifold of squeezed Bell-like states with fixed energy, we have thus performed a similar numerical optimization as what done before for the case of fixed entanglement.

\begin{figure}[b]
\includegraphics[width=6 cm]{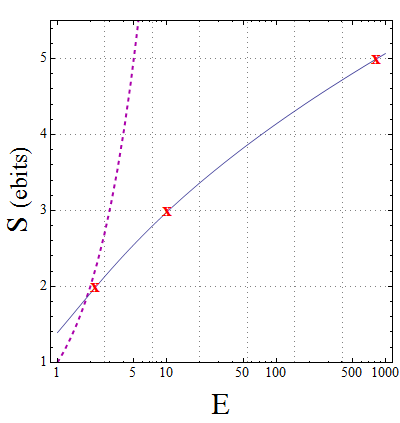}
\caption{(Color online) The dependence of the entanglement entropy $S$ of the resource states as a function of their mean energy $E$, plotted for: (a) the multiple Bell resource states for the AR scheme (dashed line) and (b) the optimal TMSV resource states for the VBK scheme. For the latter, the points that correspond to $S=2,\,3,\,5$ ebits are marked with crosses to show explicitly the need for large energies (notice the log-linear scale).}
\label{Figure 7} \end{figure}

\subsubsection{Results for squeezed states}
The teleportation of squeezed states at fixed energy yielded the same results on the optimality of the entangled resources ${\left| {{\phi _{SB}}} \right\rangle _{AB}}$ of the VBK scheme: the optimal resource state turns out to be the TMSV over the whole parameter range, yielding no non-Gaussian advantage. This observation enables us to make a neat comparison to the fixed entanglement case. In Fig.~\ref{Figure 7} we show the dependence of the entanglement entropy on the mean energy, for the optimal TMSV resource state; the points  corresponding to $S=2,\,3,\,5$ ebits are marked explicitly.  As we see, the energies $E_{\text{VBK}} = 2,\,3,\,5\,\, \rm{units}$ that we consider, correspond to entanglement entropies $1.8 \le S \le 2.5$ ebits for the TMSV state. Hence, the performances of the VBK protocol will be similar to the ones shown in Figs.~\ref{Figure1},~\ref{Figure2}, which correspond to $S=2,\, 3$ ebits respectively; the VBK scheme is thus expected to be always inferior compared to the AR scheme within this range of parameters.

We can see from Fig.~\ref{Figure 7} that an entanglement entropy of $S=5$ ebits corresponds instead to the massive mean photon number of about $833$ units for the TMSV used in the optimal VBK scheme. On the other hand, the AR scheme achieves the same entanglement with only $5$ photons and this dramatic difference is illustrated in the same figure. In fact, the AR scheme is so superior when considering energy as the resource, that even if we chose to follow the \textit{naive} interpretation described in Sec.~\ref{Comparison} and counted the photons expended in the failed teleportation attempts, we would still find that a $5$-arm scheme utilizes much less than $833$ photons as long as $\beta>1$, which would yield and endured dominance of the AR scheme over the VBK under these terms of comparison.

\begin{figure}[t]
\includegraphics[width=6.5 cm]{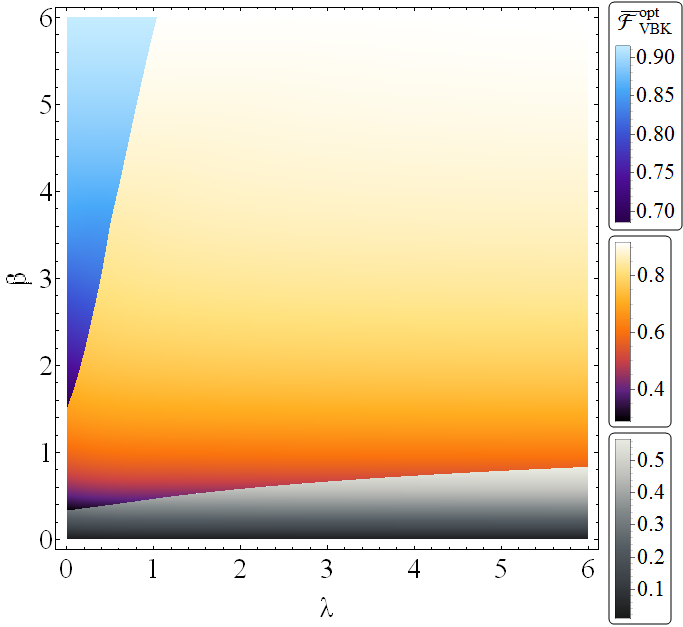}
\caption{(Color online)  Contour plot of the average teleportation fidelity $\bar  {\mathcal{F}} _{\text{VBK}}^{\text{opt}}$ for the input set of arbitrary displaced squeezed Gaussian states $|\alpha,\xi\rangle$ distributed according to the prior $p_{\lambda ,\beta }^G$, for the gain-optimized VBK scheme,   as a function of the inverse widths $\lambda$, $\beta$, at fixed mean energy of the resource states, $E=5$ units. As in Fig.~\ref{Figures456}, from top-left to bottom-right, the three shaded areas in each figure denote, respectively, the region where the VBK scheme has superior performance compared to both the AR scheme and the benchmark (sea colors online), the region where the VBK scheme is inferior to the AR one but still beats the benchmark (solar colors online) and the region where the VBK protocol yields a fidelity below the benchmark (grayscale colors online). The average fidelity of the AR protocol (not depicted) is found to always beat the benchmark for every value of the parameters $\lambda$, $\beta$.}
\label{Figure 8} \end{figure}

\subsubsection{Results for general displaced squeezed states}
We confirm once more the TMSV to be the optimal resource state for the VBK scheme, under the fixed energy constraint, when teleporting the general Gaussian set of displaced squeezed states. Adding this to the previous results, we have shown that under the restrictions of fixed energy or fixed entanglement, any non-Gaussianity within the class of squeezed Bell-like states will not give any advantage in the optimized VBK continuous variable teleportation of single-mode Gaussian states. We discussed above the relation between entanglement and energy for the optimal TMSV and showed that, for an energy of $E=5$ units, its entanglement is about $2.5$ ebits smaller than the corresponding entanglement of the resource states used in the AR scheme at the same energy.
Despite this fact however, as we see in Fig.~\ref{Figure 8}, the VBK scheme still manages to beat the AR (and the benchmarks) for small enough values of $\lambda,\, \beta$. This shows that the AR scheme is still unable to handle broad distributions, i.e. high energy inputs, when its number of branches $N$ is not big enough. For smaller energies $E=2,\,3$ units, the comparative performance of the schemes is similar to Fig.~\ref{Figure4} since at these energies the corresponding entanglement entropy is around 2 ebits for both schemes.

\section{Conclusions}\label{SecCon}

We have compared the  Vaidman, Braunstein and Kimble (VBK) continuous variable quantum teleportation protocol \cite{vaidman,brakim}, to  the recently proposed hybrid teleportation protocol of Andersen and Ralph \cite{AR}, and to the teleportation benchmarks for  general Gaussian states recently derived by Chiribella and Adesso \cite{benchnew}. We considered two classes of input single-mode ensembles, comprised of squeezed-only states and arbitrary displaced squeezed states respectively.

For the VBK protocol, non-Gaussian two-mode resources (squeezed Bell-like states \cite{Dell'Anno2007}) were considered as shared resources and optimizations were performed in order to examine any possible advantage due to non-Gaussianity of the resources for the average teleportation fidelity.
In \cite{Dell'Anno2007,Dell'Anno2010}, it was found that, under fixed squeezing of the resource state, the presence of non-Gaussianity gave significant advantage for teleportation of displaced squeezed states. These results generalized previous findings when particular non-Gaussian states such as photon-subtracted states, which are a subclass of the squeezed Bell-like states, were analyzed \cite{dega1,dega2,dega3,dega4}.

In this paper, motivated by a closer consideration of the resources involved in teleportation protocols, we adopted different terms of comparison. We compared the performance of the various schemes either at fixed entanglement entropy, or at fixed mean energy, of the shared resource states. Under these premises, we found in all considered cases that non-Gaussianity is arguably of no advantage at all: the optimal resources with a fixed entanglement or energy were consistently found to be conventional Gaussian two-mode squeezed vacuum states when the VBK teleportation protocol was considered, taking into account gain optimization \cite{Bowen2003}.

In the case of squeezed input states, we have shown that using only minimal resources, i.e. just 2 ebits of shared entanglement between the two parties, the AR scheme can successfully beat the benchmark in teleporting squeezed states while the VBK scheme, even when gain-optimized, cannot do so in a relevant parameter range. The current technological limitations  prevent us  from attaining optical squeezing larger than about $10$ dB \cite{schnabel,schnabel13}, corresponding to a maximum of $S \approx 2.77$ ebits for the VBK scheme. Even with this maximum amount of shared entanglement, the VBK scheme is unable to beat the benchmark without gain-tuning (see Fig. \ref{Figure2}) while, when gain-optimized, although it surpasses the benchmark, it still yields an inferior performance to the one of the AR scheme. The case of the fixed energy condition was even less favourable for the VBK scheme, since restricting the number of photons in the two-mode squeezed vacuum to low numbers  greatly limits the performance of the scheme. On the other hand, the AR scheme remains as much efficient for low energies since the entanglement is densely distributed over the entangled photons of the resource states, as seen in Fig.~\ref{Figure 7}.

In the case of general Gaussian input states, we saw that the AR scheme always beats the benchmark for all values of parameters  $\beta , \lambda$ of the input ensemble, while the VBK scheme is the most efficient only in teleporting coherent states (i.e. $\lambda \to 0$ and large $\beta$). For low resources, e.g. $S=2,\,3$ ebits, the AR scheme was found to perform best in teleporting broad ensembles in squeezing because of its sensitivity to the input states, beating on average the insensitive VBK scheme and the classical benchmark. However, as we reach up to $S=5$ ebits of shared entanglement, the gain-optimized VBK scheme completely dominates AR over almost all the examined region in the teleportation of general Gaussian states except for the region that corresponds to $\beta \to 0$. We should note however that this amount of entanglement is not achievable with current technology.

While the VBK scheme has traditionally been praised for its deterministic nature, which gained its historic status of an {\it unconditional} teleportation protocol (as opposed to the initial experimental realisations of discrete-variable teleportation \cite{telezei} which relied heavily on post-selection), in this case it is this feature which appears to set it at a disadvantage. It may be thus interesting to consider probabilistic alterations to the VBK scheme to see if some advantage can be recaptured. Preliminary calculations on simple conditioning strategies, such as discarding teleportation runs when Alice's quadrature measurements result in outcomes larger than a set threshold, show a minimal improvement over the deterministic VBK scheme. It thus appears that the advantage of the AR scheme does not just stem trivially from its probabilistic nature. Regardless, we dedicated considerable attention to the issue of establishing fair conditions for comparing probabilistic and deterministic schemes for teleportation of an input ensemble; we expect such a discussion to generate further independent interest in the matter.

Our analysis reveals how hybrid approaches to continuous variable quantum technology can be particularly promising with limited resources. In the case of teleportation, splitting an ensemble of Gaussian states into as few as two or three single-photon channels and performing qubit-like parallel teleportation appears effectively more efficient, even taking into account properly the nonunit probability of success, than  realizing an unconditional continuous variable teleporter consuming as much entanglement. Interestingly, a complementary hybrid approach has also very recently been demonstrated by Furusawa and coworkers, who performed deterministic teleportation of a single-photon state by a VBK implementation \cite{furunature13}. Other schemes for the near-deterministic teleportation of hybrid qubits have also been devised \cite{addedreferee}. For a review on hybrid quantum optical communication see e.g.~\cite{ibrirev}.

We note that the analysis in the present paper has focused on ideal teleportation regimes. In a real experiment, both considered schemes will be affected by unavoidable losses and imperfections, perhaps the most important ones being the noisy production of the entangled resources. In any realistic implementation, the resource states would indeed be most typically mixed nonmaximally entangled two-qubit states for the AR case, and two-mode squeezed thermal states for the VBK case. One can then still issue comparisons at fixed entanglement degree (using e.g.~the entanglement of formation) or energy, at comparable levels of state purity mirroring the current experimental facilities. These are expected to lead to the same qualitative hierarchy between the two schemes as in the case of pure resource states. Additional sources of imperfections can be considered, like lossy transmission channels in both schemes, the non-unit efficiency of the homodyne detection in the VBK scheme, the dark counts and finite detection efficiency of single-photon detectors during the Bell measurement in the AR case, etc. In this respect, the efficiency of the Bell measurement in the AR scheme is typically much lower than the efficiency of homodyne detections in optical implementations of VBK teleportation. However, this effect is typically absorbed into a lower probability of success for the AR scheme, without impacting significantly on the teleportation fidelity. Therefore, once more, we do not expect significant changes in the comparison between the two schemes and the benchmarks from the point of view of the ensemble fidelity. In short, the analyzed schemes are expected to be quite robust to common sources of imperfection. Nonetheless, we plan to complement the present investigation of the ideal regime with a forthcoming work, where all such realistic corrections will be taken into account in detail.

To our knowledge, an experiment that verifies unequivocally the use of quantum entanglement during a quantum teleportation protocol, by violating the corresponding fidelity benchmark, has yet to be performed for an ensemble of input squeezed Gaussian states with unknown squeezing (in \cite{fernnp} the input states had unknown displacement but known squeezing). In this paper we found that the hybrid AR scheme appears to be a good candidate for such a first demonstration. With the necessary technology readily available, it would be of great interest to accomplish such an experiment in the near future. In parallel, we hope this work can stimulate further research into the definition of a possibly refined teleportation protocol tailored to displaced squeezed input  states,  able to beat both the benchmarks and the AR scheme studied here, while being ideally endowed with an improved probability of success under realistic conditions.

\acknowledgments{We warmly acknowledge Giulio Chiribella for enlightening conversations on quantum teleportation protocols and benchmarks. We further acknowledge fruitful discussions with Fabrizio Illuminati and Ulrik Andersen. This research is supported by the University of Nottingham and the Foundational Questions Institute [Grant No. FQXi-RFP3-1317].

\end{document}